\documentstyle[preprint,tighten,aps,epsf,floats]{revtex}
\begin{document}
\title{
Status of experimental investigations of $\eta$-mesic nuclei}
\author{G.A.\ Sokol%
\thanks{Talk given at International Workshop "Relativistic Nuclear Physics:
 from hundreds MeV to TeV", June 26-July 1, 2000, Stara Lesna, Slovak Republic.
E-mail: gsokol@x4u.lebedev.ru}}
\address{P.N. Lebedev Physical Institute,
  Leninsky Prospect 53, Moscow 117924, Russia}
\maketitle
\begin{abstract}
Short history of ideas concerning a possible existence
of bound states of the $\eta$-meson and a nucleus is considered. 
First experiments at BNL and LAMPF on searching for these states 
are discussed. Another recent experiment using the photon
beam of the 1 GeV electron synchrotron of LPI is described.
Possible experiments on studying $\eta$-mesic nuclei
using a proton beam (at Nuclotron of Dubna) and a $\gamma$-beam
(CEBAF, JLAB) are suggested.
\end{abstract}
\normalsize
\section{Introduction}
This report has the aim to discuss of the states at studing a new objects  of
nuclear physics - $\eta$-mesic nuclei, $_{\eta}A$, a bound system of $\eta$-meson
and nucleus (fig. 1). The $\eta$-nuclei are a new kind of the atomic nuclei and
their research has the fundamental significance in studing the interaction of
the $\eta$-meson with nucleons and nucleon resonances into the nucleus.

\vspace*{1cm}
\begin{figure}[hbt]
\epsfxsize=0.6\textwidth
\centerline{\epsfbox[89 589 489 775]{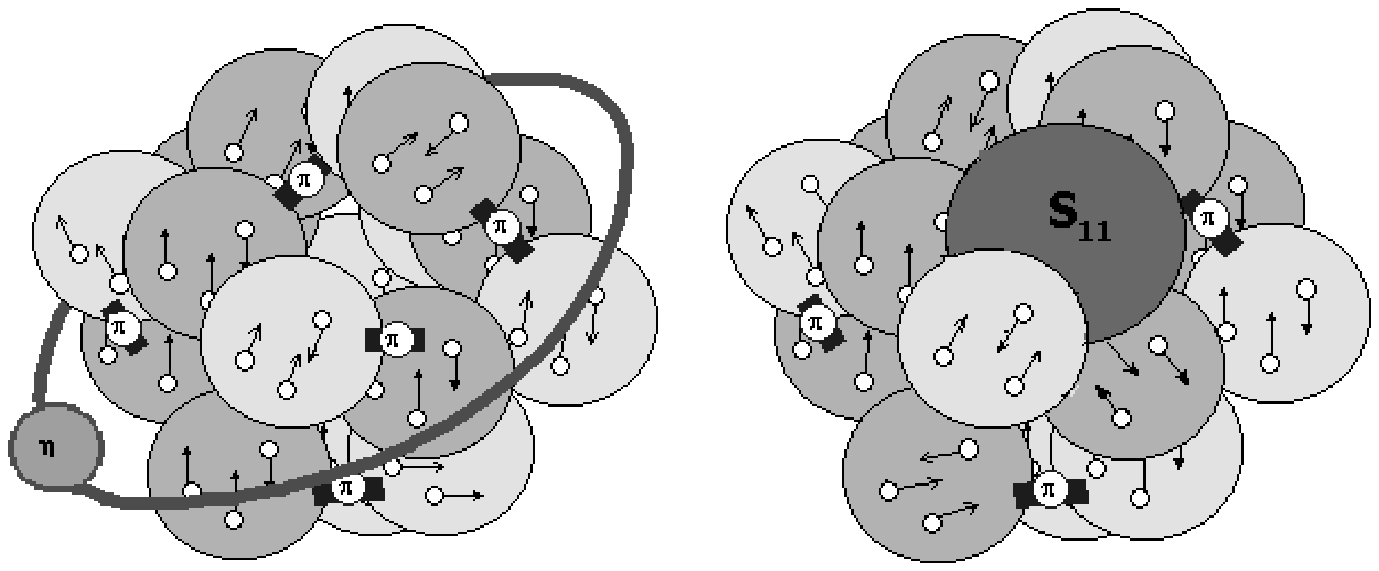}}
\caption{ $\eta $-nuclei  as $\eta $-meson in nuclei field
(left) and as $S_{11}$-resonance inside nuclei(right).}
\end{figure}

\subsection{Short history.}
Beams of the $\eta$-mesons cannot be obtained because of very short life-time
of the $\eta$-mesons $(t_{1/2} \sim 5,5 \cdot 10^{-19}$ sec). So the study of the
interactions of the $\eta$-mesons with nucleons in elementary processes of
$\eta N \to \eta N$ $\eta N \to \pi N$ type is not possible. The interaction
$\eta N$ can be investigated only in the final state of the elementary processes,
for example $\pi N \to \eta N, \gamma N \to \eta N$. \\
The $\eta$-mesic nuclei give us unique possibility for study $\eta N$-interaction
in nuclear matter. \\
The possibility of the existence of the $\eta$-meson and nucleus bound state
was firstly noted by J.C. Peng [1] in connection with the analysis of the
$\pi N \to \eta N$ reaction. As one noted, this reaction is characterized
by nonrecoil kinematics for a $\eta$-meson. Then assuming that the reaction
proceeds on a nuclear nucleon providing that: \\
\ \ \ - the $\eta N$-interaction has attractive character in a s-wave and \\
\ \ \ - the reaction time of $\eta$-meson in the nucleus is not small \\
the $\eta$-mesic nuclei can be formed.

\subsection{The scattering  length $a_{\eta N}$}.
The interaction of the $\eta$-meson in a s-wave with nucleon can be describe
by $a_{\eta N}$. The analysis of the scattering lengths of reactions $\pi N \to
\pi N; \pi N \to \pi \pi N$ and $\pi N \to \eta N$ carried out by R. Bhalerao
and L. Liu [2] has shown that scattering length $a_{\eta N}$ calculated for the
reaction $\eta N \to \eta N$ is egual to:
$$
a_{\eta N} = (0,27 + i \cdot 0,22) fm \eqno (1.1)
$$
A positive value of $Re a_{\eta N}$ means that \\
- the interaction in s-wave has attractive character and \\
- the existence of quasi-bound states of the $\eta$-meson with a nucleus for
$A \ge 11$ is possible at this value $Re a_{\eta N}$. \\
The widths of these levels were estimated by Q. Haider and L. Liu [3]. The
calculated values of width $\Gamma_g (\eta) \approx 10$ MeV mean that the
life-time of these states is appoximately 10 times larger then the "characteristic"
nucleus time $(10^{-23}$ sec), corresponding to the flight of relativistic
particle through the nucleus. \\
The values of $\Gamma_g (\eta) \simeq 30 \div 80$ MeV for the widths of the bound
states of the $\eta$-nuclei are obtained by E. Oset et. al [4]. \\
New calculations [5] of the scattering length $a_{\eta N}$ for the $\eta N \to
\eta N$ reaction were then carried out and the values of $Re a_{\eta N}$ were
found to be 3 times larger then those from paper [2]. \\
This means that the existence of the $\eta$-meson bound states with lighter nuclei
having atomic number $A = 4 \div 5$ [6] and even A=2 [7] is possible. \\
Fig. 2 shows energy dependence of $Re a_{\eta N}$ and $Im a_{\eta N}$ [5].
\begin{figure}[t]
\vspace*{-1cm}
\epsfxsize=0.7\textwidth \epsfbox[117 400 461 570]{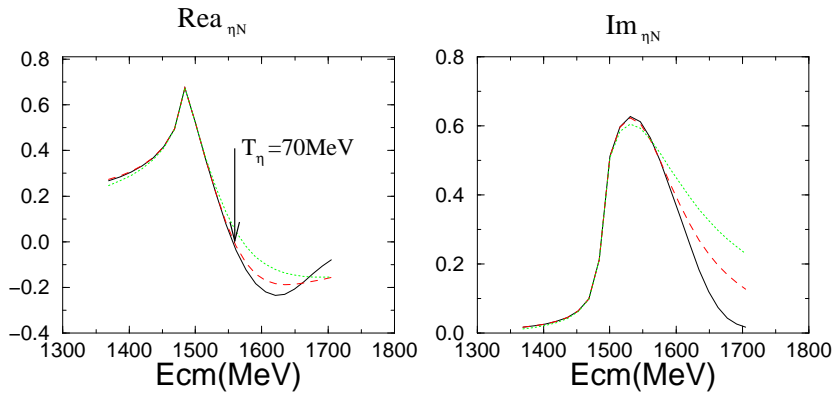}
\caption{\sl Energy dependence of Rea and Ima of
 $a_{\eta N}$ for process $\eta N \to \eta N $  [5].}
\end{figure}

\subsection{The first experiments for search of the $\eta$-nuclei.}
First experiments on direct reseach for $\eta$-nuclei were undertaked just
after calculation of $a_{\eta N}$ [2] with using $\pi^+$ beams at BNL [8] and
LAMPF [9]. The criterion of $\eta$-nuclei selection in the BNL experiment was
the observation of narrow kinematic peak $(\Gamma_g \sim 9$ MeV) in the spectrum
of proton arising in the final state of the 2-particle reaction:
$$
\pi^+ + A \to p + _{\eta}(A - 1) \eqno (1.2)
$$
The peak width $\Gamma_g (\eta)$ corresponded to the width of the $\eta$ bound
state in $_{\eta}(A - 1)$ nuclei [3]. \\
The trigger of the events was detection only one particle-proton. The
experiment was carried out on a $\pi^+$ meson beam with momentum of 800 MeV/c
ising Li, C, O and Al targets. A narrow peak, 9 MeV in width, was assumed to be
observed in the spectrum of protons for $\Theta_p = 15^0$.
However, no narrow kinematic peak was observed for any target in the spectrum
p in the expected area. The authors of paper [8] would like to note that these
negative results can be explaned by:\\
- a larger width of the expected peak than it was assumed. \\
- a smaller cross section, that it was assumed. \\
- a very bad relation: effect/background $< 1$.  \\
- a too high energy of primary $\pi^+$-mesons. \\
\ As noted in [6] $Re a_{\eta N}$ is decreased with energy and can become
negative at higher energies. This means that "attraction" in the $\eta N$ system
can be replased by "repulsion". \\
Of cousrse, all risons above are important, but in our opinion, the negative
results in [8] can be due in principle the fermi-motion of nucleons in the
nucleus. Really, a reaction of the (1.2) type at a fixed energy of $\pi$-mesons
but with the nucleon mowing in the nucleus (fermi-motion) is equivalent that on
the nucleon at rest, but with energy "spread" of $\pi$-beam.
This naturally leads to a sharp energy spread of protons detected at fixed angle
$\Theta_p$. \\
An experiment [9] was carried out on a $\pi^+$-meson beam with a momentum of
640 MeV/c and a harder trigger for event selection corresponding to the production
of $\eta$-nuclei was used: proton p was detected on coincidences with $\pi$-meson
from the decay of the $S_{11}(1535)$ resonance in the $\eta$-nucleus. As noted
in report [9], in this experiment an excess of counts was observed in the
expected area for a kinematic peak of protons. However, the experiment was not
completed, as so the results are not published. \\
Thus, first direct experiments on the discovery of the $\eta$-nuclei have not
given the expected results. At the same time, some results were obtained in the
study of reaction $p + d \to ^3He + \eta$ [10]. There interpretation requred to
use of the representations of nucleus interaction of $\eta$-meson with the
nucleus in the intermediate stage of the reaction [10]. The characteristic
properties of this reaction was:\\
- a very large cross section in a 10 MeV interval at threshold, by two orders
greater then the cross section of the reaction with $\pi^0$-meson production. \\
- practically an isotropic angular distribution of $\eta$-mesons over this energy
range (at threshold), which assumed a multiplicity of interaction of $\eta$-mesons
with nucleons inside the nucleus. \\
An indication of nuclear interaction of $\eta$-meson with the nucleus in the
intermediate stage of the reaction was also obtained in the analysis of double
charge-exchange reaction $^{18}O (\pi^+, \pi^-) ^{18}Ne$ [12], where a peak was
observed in the excitation curve at energy $E_{\pi^+} = 410$ MeV corresponding
to $\eta$-meson production threshold.

\section{First results concerning formation of $\eta$-mesic nuclei in
photoreactions.}
The problem of the $\eta$-nuclei existence for long time $(\sim 10$ years)
remained open after experiments at BNL and LAMPF with negative results. Only in
1998 in the experiment carried out on the 1 GeV electron synchrotron at Lebedev
Physical Institute the results were obtained which can be interpreted as a
direct experimental evidence for the existence of bound $\eta$-meson-nucleus
states [13].

\subsection{Method of the identification $\eta$-nuclei.}
The experiment was performed on a bremsstrahlung photon beam and correlated
$\pi^+ n$ pairs, arising from the reaction
$$
\gamma + ^{12}C \to p(n) + ^{11}_{\eta}B (^{11}_{\eta}C) \to p(n) + \pi^+ +
n + ^{10}Be (^{10}B) \to \pi^+ + n + X \eqno (2.1)
$$
have been search for. \\
The experiment was carried out at 2 energies $E_{\gamma max} = 650$ and 850 MeV,
i.e. lower and above $\eta$-meson production threshold. As noted earlier in paper [14],
the registration of $(\pi N)$-pairs and the analisis of angular and energy
characteristics can be a good criterion of production and consequent decay of
quasi-bound state of $\eta$-meson and nucleus in an intermediate stage of
reaction (2.1). \\

The $\eta$-nucleus formation in the reaction (2.1) followed by its decay is
shown schematically in fig. 3. There, the first stage of the reaction, i.e.
production of $\eta$ by photon, second stage i.e. formation of bound state
$\eta$-meson with nucleus, and the third stage, i.e. annigilation of $\eta$ and
creation of a pion, proceeds through single-nucleon interactions (either with a
proton or a neutron in the nucleus), mediated by the $S_{11}(1535)$ nucleon resonance.
Accorging to modern representations, the bound state of $\eta$-meson and nucleus
can be considered as a sequence of production and decay into $\eta N$-pairs of
the $S_{11}(1535)$ resonanse in the nucleus
$$
\eta + N \to S_{11} \to \eta + N \to S_{11} \to ... \to S_{11} \to \eta + N
\eqno (2.2)
$$
were N is nucleon, proton or neutron. \\
As result, the full averaging of the energy and angular characteristics of
pair components are arised. The decay of the $\eta$-nucleus is via the decay
of the "last" practically "termalized" $S_{11}(1535)$ resonance in this nucleus
into $\pi N$-pair.
$$
\eta N \to S_{11}(1535) \to \pi N \eqno  (2.3)
$$
Due to the Fermi motion, $\pi N$ pairs from $\eta$-nucleus decays have the
characteristic opening angle $<\Theta_{\pi N}> = 180^0$ with the width of $\simeq
25^0$. The kinetic energies are $<E_{\pi}> = 300$ MeV and $<E_N> = 100$ MeV with
the widths of $\simeq 10 \%$. It should be noted that the everage energies
$<E_{\pi}>$ and $<E_N>$ of the decay components were estimated for the $S_{11}(1535)$
resonance bound in the nucleus. Its energy is reduced to binding energy $E_g (S_{11})$
which can reach $20 \div 30$ MeV. \\
In the case when the momentum (or energy) of produced $\eta$ is high $(> 150$ MeV/c), 
the attraction between $\eta$ and the nucleus is not essential and the $\eta$-meson 
propagates freely (up to an absorprion, see Fig. 3b).
\begin{figure}[htb]
\epsfxsize=\textwidth
\epsfbox[128 650 433 730]{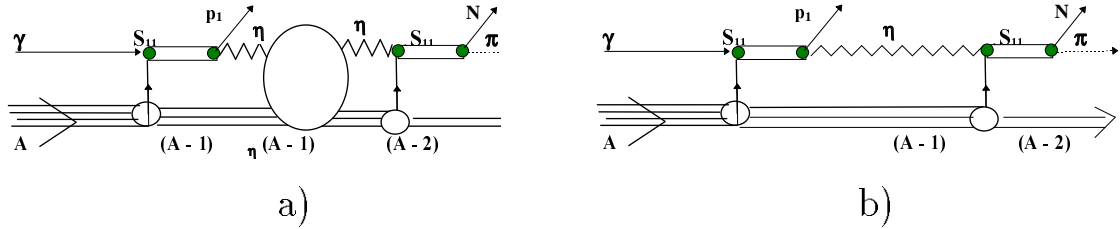}
\caption{\sl
{(left) Mechanism of formation and decay of an   $\eta$-nuclei in
photoproduction process. ($E_{\eta} \le
70$) MeV. (right) Mechanism of production and annigilation of $\eta $'s
in the nucleus. $(E_{\eta} > $70 MeV.)}}
\end{figure}

 In this case, the final 
$\pi N$-pairs also carry a high momentum and their kinematic characteristics, such 
as the opening angle $<\Theta_{\pi N}>$ and everage energies $<E_{\pi}$, $<E_N>$, 
are different from those for pairs produced through the stage of the $\eta$-nucleus 
formation. \\
The kinematics suggests photon energies $E_{\gamma} = 650-850$ Mev as the most 
suitable for creating the $\eta$-mesic nuclei. 

\subsection{Experimental set-up.}
An experimental set-up consisted of two the time of flight scintillator 
spectrometers having a time resolution of $d \tau \simeq 0,1$ ns (fig. 4). Carbon 
target $\oslash 4 \times 4$ was used. A plastic anticounter A of charged particles 
(of the $90 \%$ efficiency), placed in-front of the neutron detectors, and dE/dx 
layers, placed between "start" and "stop" detectors in the pion spectrometer, 
were used for a better identification of particles. The time of flight spectra in 
the pion and neutron spectrometers shown in fig. 4. Two-dimentional distributions 
over the time of flight particles were obtained for $(\pi^+ n)$-coincidence by 
combination their individual distribution events.                
\begin{figure}[t]
\epsfxsize=0.35\textwidth\epsfbox[114 560 305 769]{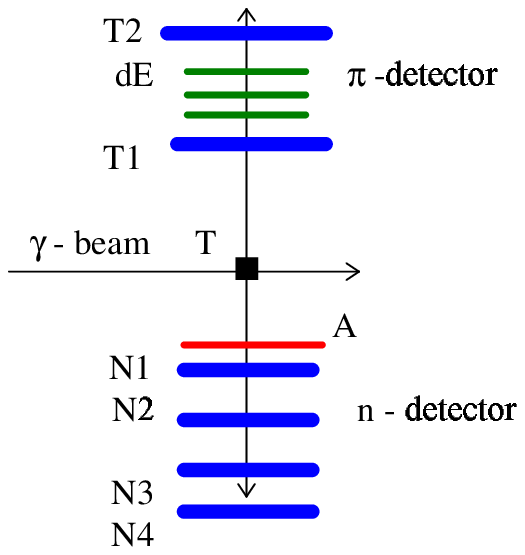}
\epsfxsize=0.30\textwidth\epsfbox[000 000 184 220]{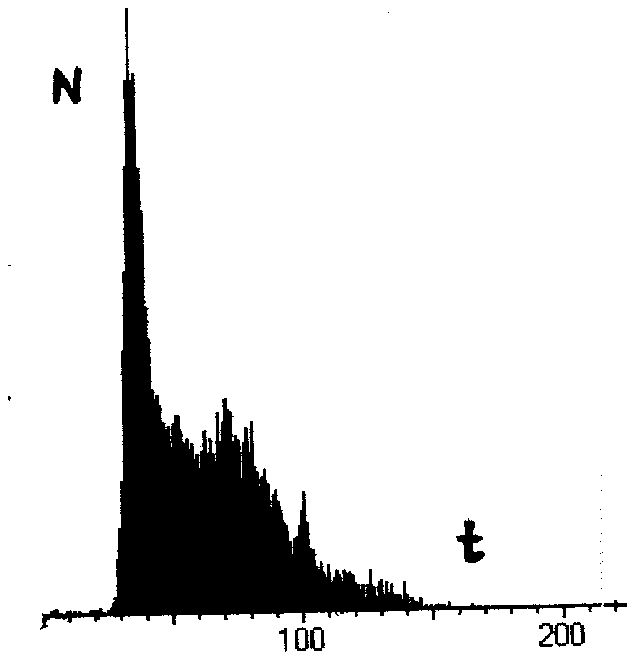}
\epsfxsize=0.30\textwidth\epsfbox[000 000 200 210]{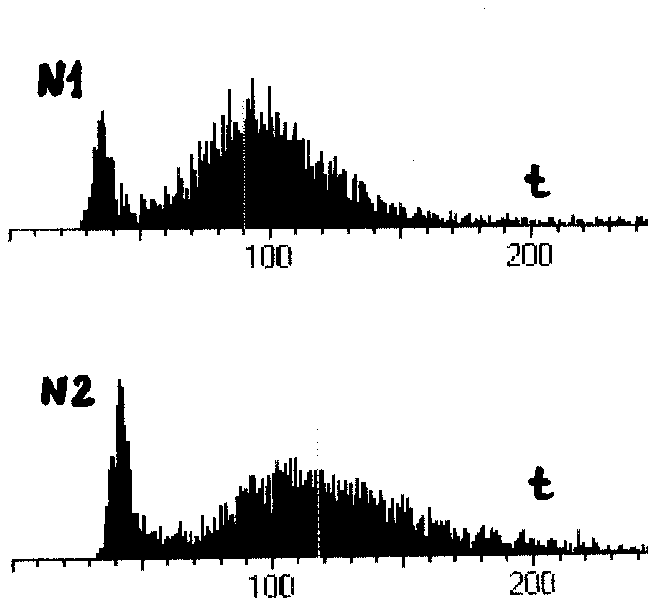}
\vspace{-2em}
\caption{Layout of the experimental setup.
Shown also time-of-flight spectra in the $\pi$ (left) and $n$
(right) spectrometers.}
\label{3}
\end{figure}
\subsection{Procedure of measurements.}                
Strategy of measurements was as follows. Two bremsstrahlung-beam energies were 
used, $E_{\gamma max} = 650$ MeV and 850 MeV, i.e. well below and well above $\eta$ 
production threshold (707 MeV on the free nucleon). The first "calibration" 
run was performed at 650 MeV with spectrometers positioned at angles $\Theta_{\pi} 
= \Theta_n = 50^0$ around the beam. In that run, this was a quasi-free 
photoproduction $\gamma p \to \pi^+ n$ which dominated, the observed yield of 
the $(\pi^+ n)$ pairs. Then, at the same "low" energy 650 MeV, the spectrometers 
were positioned at $\Theta_{\pi} = \Theta_{\eta} = 90^0$ (the "bachground" run), 
$I n$ such kinematics the quasi-free production did not contribute and the 
observed counts were presumably dominated by double pion production. At last, 
the third run (the "effect + background") was performed at the same $90^0/90^0$ 
position, however with the higher photon beam energy of 850 MeV, at with $\eta$ 
mesons are produced too. 

\subsection{Handling of raw results.}
In accordance with measured velosities of particles detected by the spectrometers 
all candidates to the $(\pi^+ n)$-events were separated into three classes: 
fast-fast (FF), fast-slow (FS), and slow-slow (SS) events. The FF events mostly 
correspond to $\pi^0 \pi^0$ production with results in hitting detectors by 
photons or $e^+/e^-$. The FS events mostly emerge from $(\pi \pi) + (\pi^+ n)$-pairs. 
The SS events arised from $(\pi \pi)$ pairs. Comparing yields and time spectra in 
these runs we have found a clear excess of the FS events which appeared when the 
photon energy exceeded $\eta$ production threshold. (see [13] for more details). 
The raw experimental spectrum over the  particles velocities had unphysical region 
with $\beta_i > 1$. So happened the velocities $\beta_i = L_i/t_i$ are subject to 
fluctuations stremming from errors $\delta t_i$ and $\delta L_i$ in the time-of 
flight $t_i$ and the flight base $L_i$. Such fluctuations are cleary seen in the 
case of the ultra-relativistic FF events. Therefore, an experimental 
$\beta$-resolution of the set-up can be directly inferred from the FF events. 
Then, using this information and applying an inverse-problem statistical 
method discribed in Ref [15], one can unfold the experimental spectrum, obtain 
a smooth velocity distribution on the physical region $\beta_i \le 1$. (fig. 5), 
and eventually find a distribution of the particle's kinetic energies 
\begin{figure}[htb]
\leavevmode
\epsfxsize=15cm
\centerline{\epsfbox[37 260 548 582]{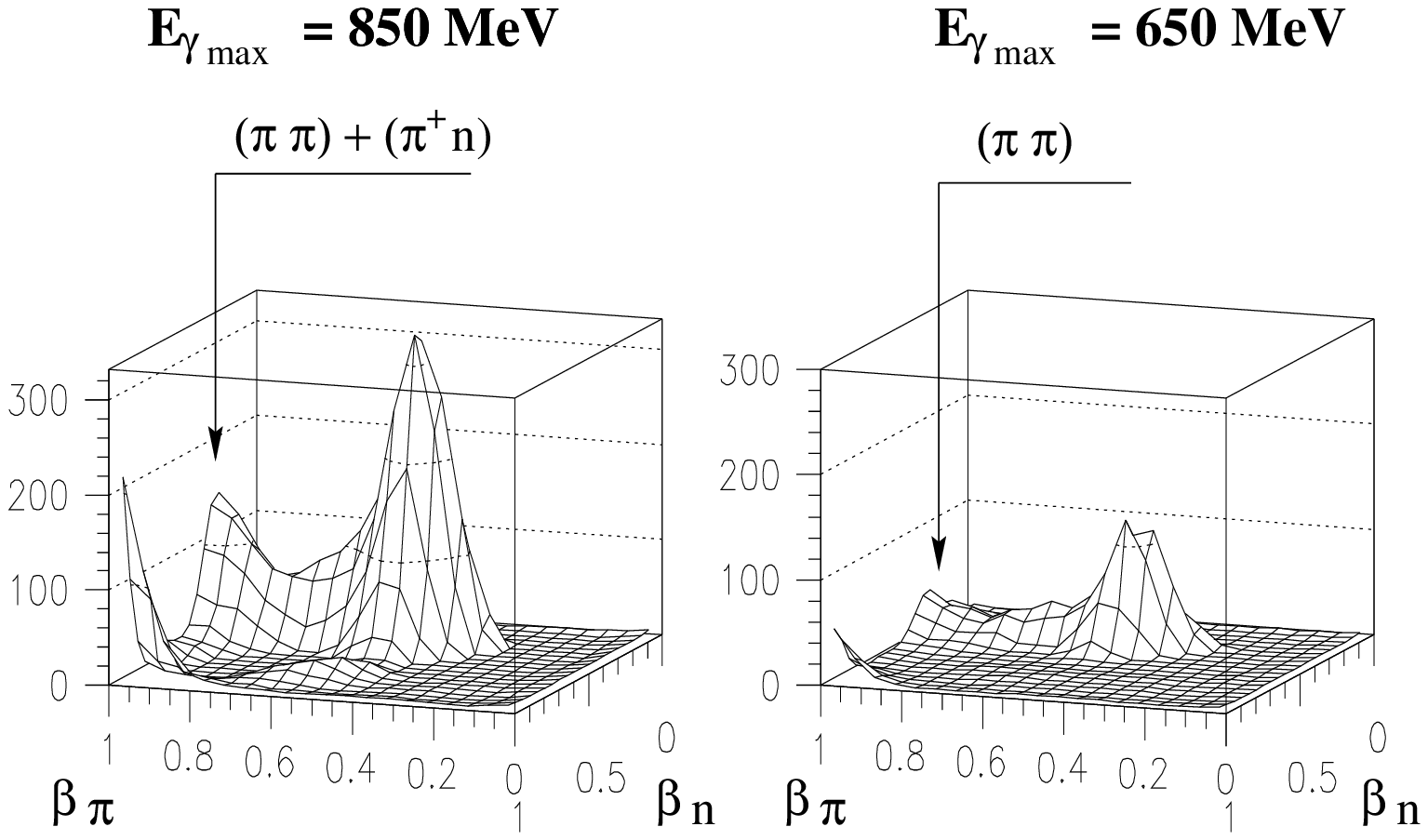}}
\caption{Corrected two-dimensional distributions over the velocities ($\beta $)
for the $(\pi^+n)$-events with end-point energy of the bremsstrahlung  spectrum
 $E_{\gamma max}$ = 650 and 850 MeV.}
\end{figure}

\begin{figure}[th]
\unitlength=1mm
\begin{picture}(100,55)(0,0)
\put(40,48){$E_{\gamma \rm max}=850$ MeV}
\put(90,48){$E_{\gamma \rm max}=650$ MeV}
\centerline{\epsfxsize=4.5cm\epsfbox[157 322 443 613]{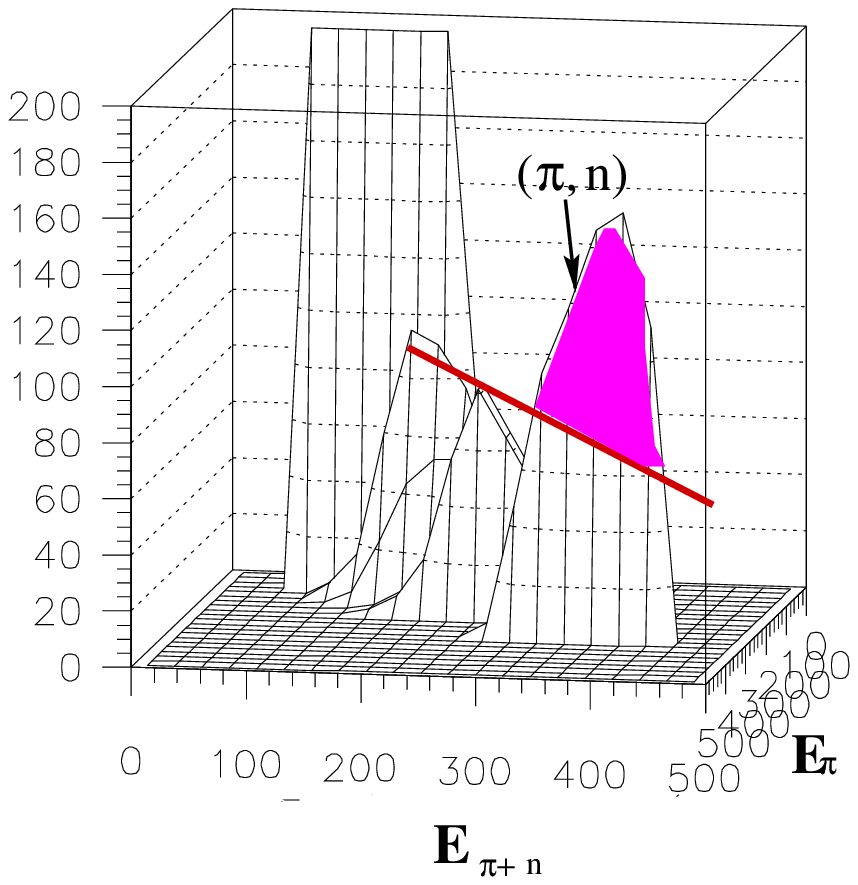}
\epsfxsize=4cm\epsfbox[16 325 304 670]{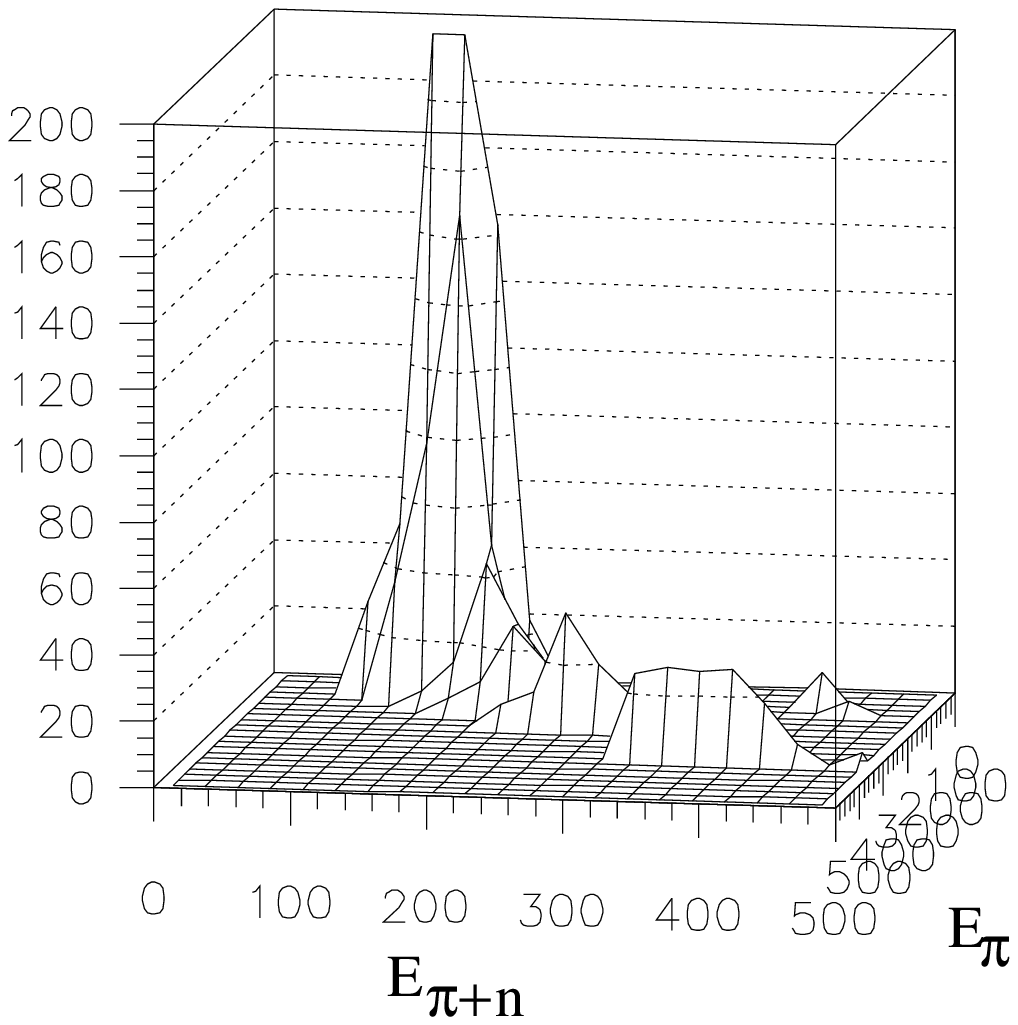}}
\end{picture}
\caption{Distribution over the total kinetic energy of the $(\pi^+n)$ pairs
for the ``effect$+$background" run (the left panel) and for the
``fon" run (the right panel) obtained after unfolding raw spectra.}
\label{fig:Etot-2dim}
\end{figure}
\begin{figure}[htb]
\epsfxsize=0.4\textwidth
\centerline{\epsfbox{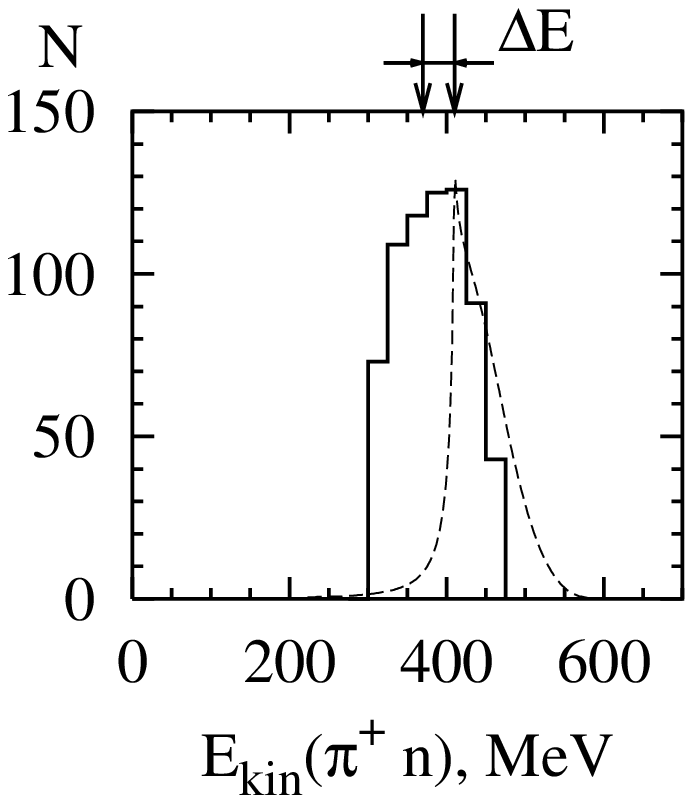}}
\caption{Distribution over the total kinetic energy of the $(\pi^+n)$ pairs
after subtraction of the background. An arrow indicates the threshold of
408 MeV (see in the text). For a comparison, a product of free-particle
cross sections of $\gamma N \to \eta N$ and $\eta N \to \pi N$ is shown
with the dashed line (in arbitrary units).}
\label{fig:Etot-1dim}
\end{figure}

$E_i = M_i [(1 - \beta^2_i)^{-1/2} - 1]$. Finding $E_i$, we introduced corrections 
related with every energy losses of particles in absorbers and in the detector 
matter. It is worth to say that the number of the $(\pi^+ n)$ FS events visibly 
increases when the photon beam energy becomes sufficient for producing $\eta$ 
mesons. 
\subsection{Results.}
Of the most interest is the distribution of the $(\pi^+ n)$-events over their 
total energy $E_{tot} = E_n + E_{\pi}$, because creation and decay of $\eta$-mesic 
nuclei is expected to produce a relatively narrow peak in $E_{tot}$ (see [13, 16]. 
Such a peak was indeed observed.

 At fig. 6 we see an excess of the FS events 
appears when the photon energy exceeds the $\eta$-production threshold. These 
$(\pi^+ n)$ pairs arise from creation and decay of captured bound $\eta$ in the 
nucleus, i.e. they arise throngh the stage of formation of an $\eta$-mesic nucleus. 
The second important results is a "shift" of the position of $S_{11}(1535)$ 
resonance  decayed into nuclei.

 On fig. 7 we have a 1-dimensional energy 
distribution of the $(\pi^+ n)$ events presumably coming from bound $\eta$ 
decaing in the nuclei. The experimental width of this distribution is about 100 
MeV including the apparatus resolution. Its center lies by $\Delta E = 40$ MeV 
below the energy excess $m_{\eta} - m_{\pi} = 408$ Mev  in reaction $\eta N \to 
\pi N$, and it is well below the position of the $S_{11}(1535)$ resonance too. \\
Up to effects of binding of protons annihilated in the decay subprocess $\eta p \to 
\pi^+ n$, the value $\Delta E$ characterizes the binding energy of $\eta$ in 
the nucleus. The width of that peak is determined both by the width of the 
$\eta$-bound state and by the Fermi motion. 

\subsection{Conclusions.}
The first results which gives the evidence for existence of the bound state of 
$\eta$-meson and nucleus in intermediate stage photomesonic processees are 
obtained at LPI [13]. 

\section{Study for the $\eta$-mesic nuclei in {\normalsize p}A-collisions at JINR NUCLOTRON 
(proposal)}
\subsection{Experimental conditions.}
The use of intensive monochromatic beams of protons with energies of some GeV 
may look promising to study $\eta$-mesic nuclei. The possibility of the $\eta$-nuclei 
production in the reaction 
$$
p_0 + A \to p_1 + p_2 + _{\eta}(A - 1) \to p_1 + p_2 + p_3 + \pi^- + X \ \ \ 
\eqno  (3.1)
$$
for light nuclei with $A \ge 12 (^{12}C, ^{14}N, ^{16}O)$ is considered in this 
suggestion. \\

Fig. 8(a) presents a diagram of process describing the stage of $\eta$-meson 
production in nucleus with formation of the bound state of $\eta$-meson and nucleus 
and at last the stage of the $\eta$-nucleus decay. The diagram corresponding 
to process where the $\eta$-mesic nucleus is not production is shown in fig. 8(b). 
One can assume that the kinetic energy of the $\eta$-meson is rather great 
$(T_{\eta} > 70$ MeV) in this case and there is not attraction between $\eta$-meson 
and nucleon [6]. \\
The production cross sections of $\eta$-meson in elementary processes $\pi p \to 
\eta p$ and $p p \to \eta p p$ are practicaly similar. However, the advantage 
of p-beams consist in their much greater intensity (by 2-3 orders) in comparising 
with $\pi$-meson beams (the latter are secondary beams of proton accelerators). 
This circumstance is important as the production cross sections of $\eta$-nuclei 
in $p A$-collisions are expected at a level of some tens microbarns. \\
The event selections of $\eta$-mesic nuclei production in $p A$-collisions are assumed 
to be made by registrating 4 particles: $p_1$ and $p_2$ protons produced in the 
1-st stage of the reaction (3.1) and $(\pi^-, p_3)$ pairs from the decay of the 
$\eta$-mesic nucleus (fig. 8). To separate from the process shown by the diagram 
in fig. 8b should be detected the 4 particles in the corresponding energy and 
angular intervals. \\
For a bound state of $\eta$-meson and nucleus the angular distribution of the 
$(\pi^-, p_3)$ components from the $S_{11}(1535)$ resonance decay is characterized 
by isotropy at an everagy angle of $< \Theta_{\pi p} > \simeq 180^0$ and an 
increased yield of such pairs a 3-5 times fold in comparison with the case without 
$\eta A$-interaction.\\ 
\vspace{1cm}
\begin{figure}[thb]
\epsfxsize=\textwidth
\epsfbox[76 656 530 775]{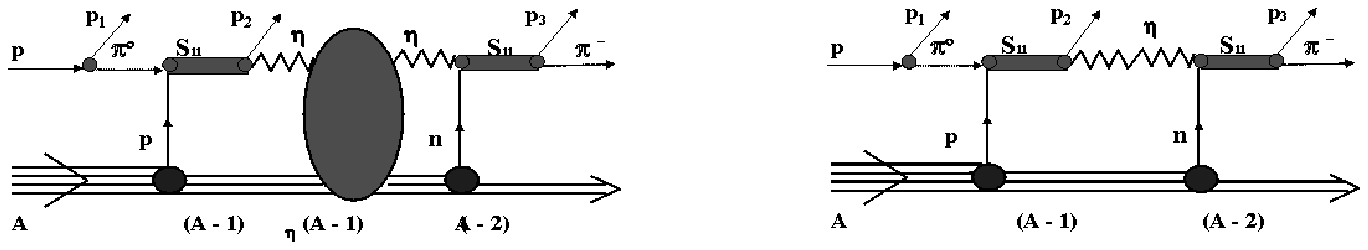}
\caption{\sl
{(left) Mechanism of formation and decay of an   $\eta$-nuclei in pA-collisions. ($E_{\eta} \le
70$) MeV. (right) Mechanism of production and annigilation of $\eta $'s
in the nucleus. $(E_{\eta} > $70 MeV.)}}
\end{figure}

\subsection{Experimental setup.}
The experiment is supposed to be performed on an internal beam of protons at the 
JINR NUCLOTRON. \\
The use of the internal beam affords unique opportunities to study processes with 
small cross sections through the full interaction of the beam with a "wire" 
target so the total intensity can reach $\sim 10^{14}$ p/hours and a low nouse 
level since the background particles are not practically multiplied into the 
target. \\

The scheme of the experimental setup and its deposition on the ring of the 
accelerator is shown in fig. 9. According to the shosen algorithm of event 
selection of the $\eta$-nuclei production in $p A$-collisions, the setup consists 
of 3 types detectors. All the detectors are made of the plastic scintillation  
counters. This is determined by the necessity of the realization of fast 
coincidences and the measurement of the time-of-flight of particles in the 
picosecond range. The detectors of $p_1$ and $p_2$ represents an assembly of 
counters W1-W8 placed arround the axis of an incident beam at an angle 
$<\Theta>$ of $\sim 10^0$. \\
The function of detectors $p_1$ and $p_2$ is to measure the coordinate of the 
detected protons, angle $\Theta(p)$ and time-of-flight. The detector used to 
register the proton $p_3$ is a scintillation telescope of the $\Delta E - E$ 
tipe. The detector of $\pi^-$-mesons is scintillation telescope of 2 counters 
TC1 and TC2. The time resolution of all detectors is about 150-170 psec. The 
first counter of the $\pi^-$-spectrometer is used as a "start" counter for all 
the time-of-flight system. 

\begin{figure}[htb]
\epsfxsize=0.9\textwidth
\centerline{\epsfbox[48 535 570 788]{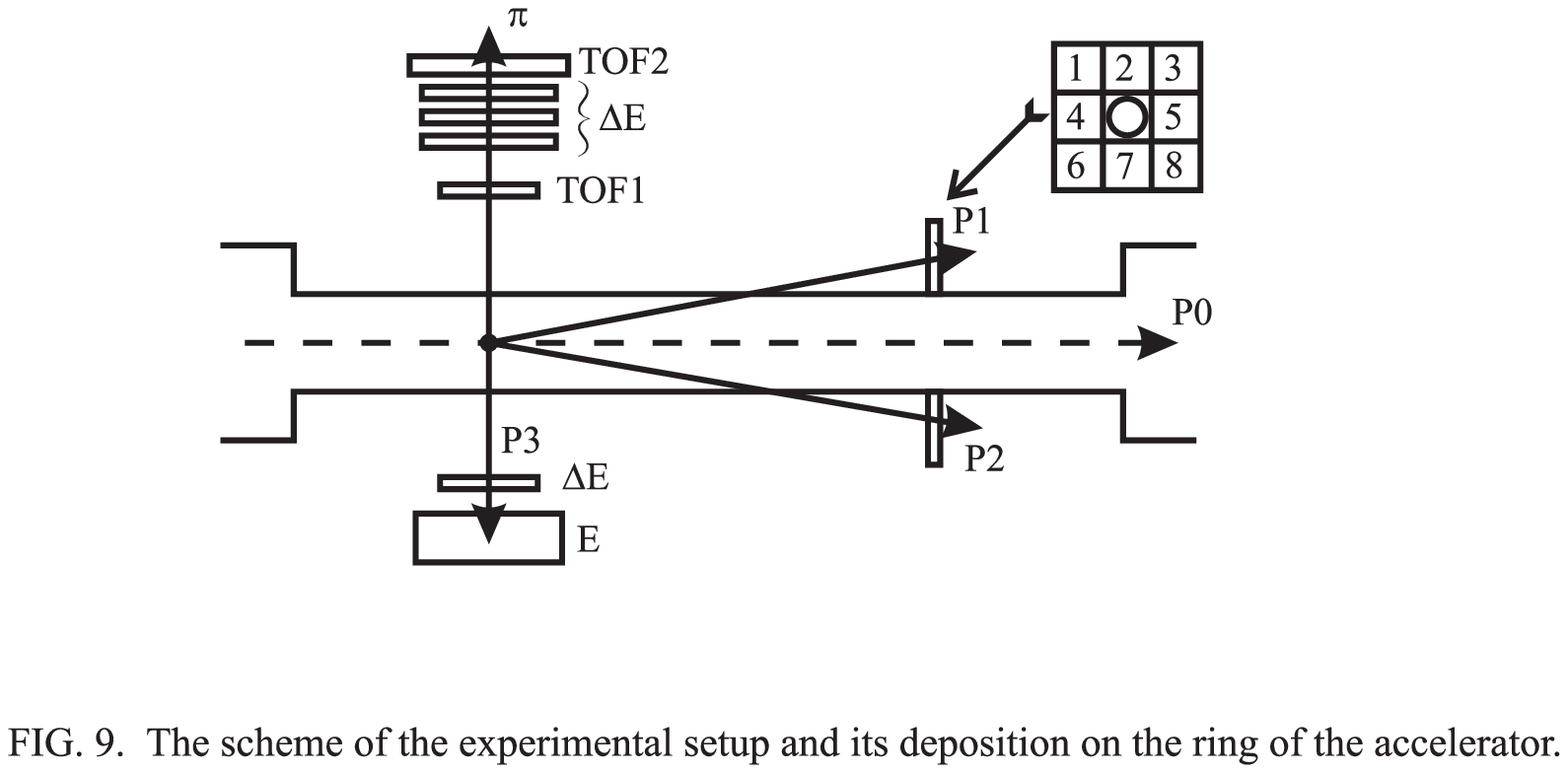}}
\label{fig.9.}
\end{figure}

\subsection{Estimates of the $\Delta Y (p_1, p_2, p_3, \pi^-)$ yield.}
As the production of the $\eta$-nuclei is preferable for slow $\eta$-mesons 
$(E_{\eta} \le 70$ MeV) the reaction yield is determined by threshold values 
of the cross sections for elementary processes of the $\eta$-meson production. 
A favourable circumstance is that production cross sections are rather significant. 
It is due to the disposition of the $\eta$-meson production threshold inside 
the energy region of the $S_{11}(1535)$ resonance, naving a large width of 
$\Gamma \sim 150$ MeV in an elementary $p p \to p S_{11}(1535) \to p p \eta$ 
reaction. The total production cross section of $\eta$-meson near threshold is 
equal to [17, 18] : 
$$
\sigma_t(p p \to \eta p p) \simeq 100 \mu kb (10^{-28 cm^2}) \eqno  (3.3.1)
$$
The yield of the 4 multiple coincidences $\Delta Y (p_1, p_2, p_3, \pi^-)$ for 
50 $\mu m ^{12}C$ target wire can be estimated from the following relation: 
$$
\Delta Y (p_1, p_2, p_3, \pi^-) = \sigma_t (p ^{12}C \to p_1 p_2\ ^{11}_{\eta}B) 
\cdot N_n \cdot N_p \cdot Br (\pi, N) \cdot \xi \cdot \Omega_{\pi} \cdot 
f(\Omega_p/\Omega_{\pi}) \cdot \eta_{\pi} \cdot \eta_p \cdot \chi p_1 \cdot 
\chi p_2 \eqno (3.3.2)
$$
In this expression: \\
$\sigma_t (p ^{12}C \to p_1 p_2\ ^{11}_{\eta}B)$ is taken as part of $\sigma_t$ 
for the $\eta$-meson production on the $^{12}C$ nucleus. This part is the same 
as for photoreactions, i.e. $\sim 5 \%$. Then the production cross section of 
the $\eta$-nuclei in the interaction of p-beam with the $^{12}C$ nucleus is equal:\\ 
$
\bullet \sigma_t (p ^{12}C \to p_1 p_2\  ^{11}_{\eta}B) = [6 \cdot \sigma (p p \to \eta p p ) 
\cdot F (^{11}B)] \cdot 0,05 = 3 \cdot 10^{-3} mb = 3 \mu kb 
$\\
where $F(^{11}B)$ is the form factor of the nucleus $^{11}B$, assumed to be $\sim 0,1$.\\ 
$
\bullet N_n = \frac {N_a}{A} \cdot \rho (g/cm^3) \cdot \Delta x(cm) = \frac {6 \cdot 10^{23} 
\cdot 1,7 \cdot 5 \cdot 10^{-3}}{12} = 4,25 \cdot 10^{20} nucleus/cm^3  
$\\
$
\bullet N_p = I_p \cdot n \cdot K = 10^{11} \cdot 800 \cdot 2 = 1,6 \cdot 10^{14}  
p/hour 
$\\
Where n=800 -number of acceleration cycles per hour, k=2 is the everagy 
multiplisity of the passages of the proton beam through the target. \\
$\bullet \ \Omega_{\pi} = \left ( \frac {50}{150} \right)^2 \frac{1}{4 \pi} = 
9 \cdot 10^{-3}$ \\
$\bullet \ Br (\pi N) = 0,5$ \\
$\bullet \ \xi = 1/3$ - the part $(\pi^- p)$ decay $S_{11}$ of all kind decay. \\
$\bullet \ f(\Omega_p/\Omega_{\pi})$ - is the correlation function = 0,2 \\
$\bullet \ \eta_{\pi} = 0.8$ - effeciency of $\pi$-detection \\
$\bullet \ \eta_p = 0,8$ - efficiency of p-detection \\
$\bullet \ \chi (p_1)$ and $\chi (p_2)$ are the geometric factors describing  the 
fraction over angle interval of $\Delta \Theta = 10 \pm 5^0$. \\
This fraction is registered by the $p_1$ and $p_2$ detectors of the total 
number of $p_1$ and $p_2$ produced in the $p p \to \eta p p$ reaction. The 
kinematic calculations give the following values: 
$$
\chi (p_1) = 0,185; \chi (p_2) = 0,233
$$
Substituling these numerical values to formula (3.3.2) we obtained: 
$$
\Delta Y (p_1 p_2 p_3 \pi^-) \cong 4 \frac {events}{hour} \simeq 100 events/day 
$$
The value of the expected yield can be considered as rather high under the 
condition of the small background level. 

\subsection{Conclusion.}
The research of the $\eta$-mesic nuclei in pA-collisions can be interest from 
point of view of development of the presentations about interaction adrons 
$(\eta$-meson) with nuclear matter and a reseach of the new kind of atomic 
nucleus - $\eta$-mesic nuclei. In these reserch one can have possibility to 
formation two-$\eta$-mesic nuclei and mesic nuclei with ofher mesons $(\rho, 
\omega, \varphi)$. 

\section{Search for a few-body $\eta$-mesic nuclei at JLab \\
(CEBAF)\\
(Suggestion for experiment)}
\subsection{Experimental conditions.}
Perfomance of study for $\eta$-mesic nuclei at CEBAF can be very favourable 
counsequently high intensity and continuous e-beam of accelerator. The 
experiment can be performed at bremsstrahlung photons in region of energy 
$E_{\gamma max} = 600 \div 1000$ Mev. \\
\vspace{1cm}
\begin{figure}[thb]
\epsfxsize=\textwidth
\epsfbox[25 430 570 780]{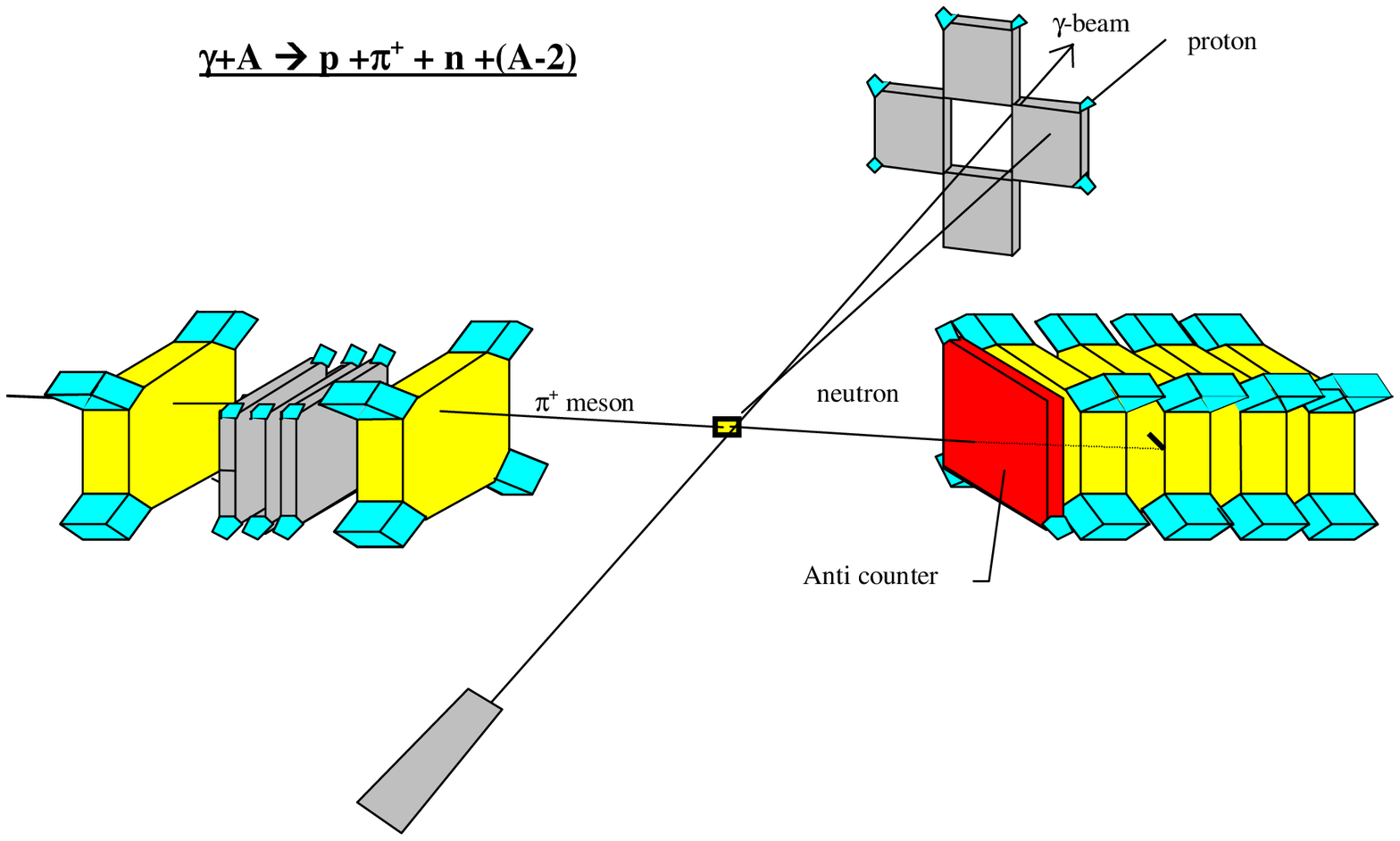}
\caption{\sl Scheme of the experimental setup.}
\end{figure}

The main task of experiment may be to measure of energy and A-dependence cross 
section for photoproductuon of light $\eta$-mesic nuclei (up A=12 to A=3). \\
The method of $\eta$-nuclei identification consist of the detection 3-particles: 
one particle, from first stage of reaction (see fig.10) (n or p) and two-particles 
from second stage -decay of $\eta$-mesic nuclei: $(\pi^+ n)$ or $(\pi^- p)$-pairs. 
Detection of 3-ple coincidence is guarantee of selection of events connecting 
with formation $\eta$-mesic nuclei. The experimental set-up must have 3 type 
scintillation time-of-flight spectrometers (Fig. 10). The yeild of 3-ple 
councedence events can be to compose about 10 events/hour. One can receive the 
experimental results about energy bounding $E_g S_{11}(1535)$ resonance in light 
nuclei. \\

One of interested results can be evidence of existence lightest $\eta$-mesic 
nuclei $(^3_{\eta}H, ^3_{\eta}He, ^4_{\eta}He)$. 

\section{Conclusions.}
Study of the $\eta$-mesic nuclei is new very interesting field of nuclear 
physics and particle physics. One can be received a new information for 
interaction of the $\eta$-meson 
with nucleon and nucleon resonance in the nuclear matter. The measurement of the 
bonding energy $E_g (\eta)$ and $E_g (S_{11})$ can be used in chiral symmetry 
theories in problem for origion of the elementary particle masses.       \\
This work was supported by RFBR grant 99-02-18224. \\
Author thank S.I. Nikolsky and E.I. Tamm for attention for work, my colleagues 
T.A. Aibergenov, Y.I. Krutov, A.I. Lvov, L.N. Pavlyuchenko, S.S. Sidorin for 
active participance in experimental work and handling results, and L.I. Goryacheva 
for help in preparing this report.

\end{document}